\documentclass[aip,jcp,amsmath,amssymb,reprint]{revtex4-2}
\usepackage{color}
\usepackage[dvipsnames,svgnames,table]{xcolor}
\usepackage[colorlinks=true,linkcolor=blue,urlcolor=black,citecolor=blue]{hyperref}
\usepackage{graphicx}
\usepackage{bmpsize}
\usepackage{dcolumn}
\usepackage{bm}
\usepackage{xcolor}
\usepackage{enumerate}
\usepackage{booktabs}
\usepackage{threeparttable}
\usepackage{amsmath}
\usepackage[utf8]{inputenc}
\usepackage[T1]{fontenc}
\usepackage{mathptmx}
\usepackage{etoolbox}
\usepackage{natbib}
\usepackage[normalem]{ulem}

\makeatletter
\def\@email#1#2{
 \endgroup
 \patchcmd{\titleblock@produce}
  {\frontmatter@RRAPformat}
  {\frontmatter@RRAPformat{\produce@RRAP{*#1\href{mailto:#2}{#2}}}\frontmatter@RRAPformat}
  {}{}
}
\makeatother

\begin{document}
\title{Latent space design of interatomic potentials}
\author{Susan R. Atlas}
\email{susier@unm.edu} 
\affiliation{Departments of Chemistry \& Chemical Biology and Physics \& Astronomy, and Center for Quantum Information and Control, Quantum New Mexico Institute, University of New Mexico, Albuquerque, NM USA}

\date{\today}

\begin{abstract}
The advent of neural-network-based deep learning techniques has led to the emergence of increasingly sophisticated numerical interatomic potentials, including graph neural networks and large language-motivated foundation models. Parameterized to reproduce large, precomputed quantum mechanical training datasets for molecules and materials, models can be fine-tuned for greater accuracy on specific problems. Despite notable successes, machine learning models of potentials still 
face intrinsic challenges due to the combinatoric complexity of the underlying quantum chemical interactions, the existence of as-yet-undiscovered but potentially relevant bonding motifs absent from training datasets, and the need for post-prediction interpretability analysis.
Drawing inspiration from autoencoder methods, we propose a constructive approach to interatomic potential design. In standard autoencoder architectures, a machine learning (ML) model self-organizes numerical training data in an unsupervised manner to discover underlying patterns and construct a compressed representation or {\it latent space} embedding model of the data, which is then used for prediction and inference. In the present work, we describe how latent space patterns and associated quantum embeddings can be constructed using first-principles methods based on theorems of density functional theory (DFT) and known, analytic constraints. This enables a parsimonious, physics-based representation of energies and densities, formally coupling the electronic and atomic length scales through the electron density, and linking ground, excited, and charge-transfer states of the interacting atoms. We describe the complete set of latent space components providing the foundation for a recently-proposed ensemble charge-transfer potential, and discuss opportunities for synergy in the design and explainability of contemporary machine-learned interatomic potentials. 
\end{abstract}

\maketitle

\section{\label{sec:Introduction} Introduction}
Interatomic potentials define the forces governing the motion of atoms in molecular dynamics (MD) simulations of molecules and materials.
They serve as classical proxies for the underlying electronic structure---the quantum mechanics---of interacting atoms. The design and parameterization of a potential model plays a critical role in accurately describing phenomena as diverse as ligand-protein binding, reactive chemistry, and
dislocation dynamics in mesoscale materials.
 
Ideally, the design of an interatomic potential should be independent of the structural organization, scale, and phase of the system under study.
From a chemical standpoint, the representation should be capable of describing bond formation and breaking, charge excitation and charge transfer, defect dynamics, and arbitrary combinations of diverse elements from the Periodic Table. These physical and chemical \textit{desiderata} 
imply two key requirements for successful classical potential design: (i) a representation capable of characterizing the instantaneous quantum mechanical environment or {\it chemical context} of an atom; and (ii) a mechanism for systematically coupling the electronic and atomistic length scales of the system over the course of a dynamical simulation. Requirement (ii) includes the need to describe nonadiabatic quantum effects arising from the breakdown of the Born-Oppenheimer approximation due to the presence of conical intersections and avoided crossings,\cite{yarkony2012nonadiabatic} and the electronic coupling of ground and excited state potential energy surfaces\cite{dreuw2005single} over the course of a dynamical simulation.\cite{tully1971trajectory,tully1990,wang2016recent,curchod2018ab,freixas2021nonadiabatic}
 
Contemporary machine learning interaction potential (MLIP) models\cite{jacobs2025practical} such as those based on graph neural networks\cite{gilmer2017neural,duval2023hitchhiker} tackle this problem by utilizing 
 datasets of quantum-mechanically-computed energies\cite{fn3} to sample and fit a broad range of exemplar atomic structures and interactions, effectively learning the quantum mechanical behavior inherent within a given training dataset.
 However, ML models are subject to the so-called ``curse of dimensionality''
 in scaling to large system sizes and increasing numbers of distinct chemical elements. In some cases, it is possible to reduce the size of a large configurational sampling volume through the use of constraints, for example
 by considering a limited number of crystallographic phases in a material,\cite{faber2016machine,roy2020machine} 
or in the protein structure prediction problem, utilizing the biological constraint of a strict sequence of amino acids linked by peptide bonds, and incorporating evolutionary information and homology modeling. \cite{jumper2021highly,abramson2024accurate,elnaggar2021prottrans}
Unfortunately, for the general problem of constructing an interatomic potential for arbitrary interacting atoms, such reductions are not possible, and increasing the number of distinct chemical elements gives rise to exponential scaling in the number of structural configurations that need to be considered.
 For sufficiently large and chemically-diverse systems, it becomes impossible to characterize all 
chemically-reasonable local bonding environments for inclusion in a quantum mechanical training set (and perhaps some unexpected, ``unreasonable" ones as well.)
 
 While the emerging generation of ML foundation models\cite{yuan2025foundation} for constructing interatomic potentials based on quantum mechanics---recent examples include the atomic cluster expansion-based MACE\cite{batatia2022mace,kovacs2023evaluation,batatia2025foundation} and Unified Model Architecture (UMA)\cite{wood2025family}---are intended for downstream tuning and specialization to simulate particular classes of molecules and materials, there is a risk of inconsistent predictions when subdomain boundaries are crossed in complex systems. Challenging cases include problems that combine subsets of small molecules, biomolecules, nanoscale materials, surfaces, and bulk or defected materials, or where the relevant subdomain has been undersampled in the training set.\cite{focassio2024performance}  Other issues
faced by MLIPs include their reliance on density functional theory (DFT) calculations based
on approximate models of electron correlation to construct
training datasets; incomplete sampling of chemically-relevant structures;  and a complex mathematical formalism emphasizing chemical accuracy at the risk of obscuring chemical interpretation of the resulting dynamics (the ``explainability" issue in deep learning.)
 
The goal of the present work is to describe a possible way out of this dilemma, by encapsulating pre-existing, formally-constituted quantum mechanical knowledge to define a chemically-motivated \textit{constructive latent space} representation, and demonstrate how this representation can be used as the starting point for designing physics-based interatomic postentials.  
The latent space approach avoids the requirement that a machine learning (ML) process bear the entire burden of decoding and mathematically modeling the quantum mechanics of electron correlation.
In contrast to the numerical features and descriptors of traditional ML models, the latent space \textit{patterns} correspond to electronic structure design elements. They can be understood as the analogs, at the quantum scale, of design elements used in fitting empirical potentials based on canonical arrangements of atoms: secondary structure in a protein, crystal structures defined by space group symmetries, Hookean springs to describe covalent bonds, sets of molecular conformers, and defects such as grain boundaries, dislocations, and vacancies in materials. 
 
 Unlike the situation at the atomic-scale, however, the appropriate design elements for constructing quantum-informed interatomic potentials are not  intuitive. This is due to the inherent complexity of the electron correlation problem:
 identifying a set of quantum-scale design elements enabling the construction of a perfectly accurate interatomic potential would  be tantamount to solving the electron correlation problem embodied in the wavefunction $\Psi$ for the entire system.
 
 The question, then, is whether it is possible to construct a quantum-level latent space representation capable of describing arbitrary interactions among atoms, without requiring {\it a priori} imposition of symmetries, structural constraints, or excessive numbers of fitting parameters. The purpose of the present work is to propose a step toward addressing this challenge. The approach described here was originally proposed as an extension of the venerable embedded-atom method,\cite{daw1983,daw1989} to account for charge distortion and charge transfer in complex interacting systems.\cite{muralidharan2007ED,valone2014,amokwao2020,Atlas2021,baxter2026,samuels2025,samuels2026information,amokwao2026} The resulting potential model, termed the {\it ensemble charge-transfer embedded atom method} (ECT-EAM), defines a general interatomic potential model in its own right.\cite{Atlas2021}  The latent space patterns consist of isolated-atom quantities that can be pre-computed,
 so that the only parameters appearing in the model are ensemble weights that self-adjust over the course of a simulation to maintain an instantaneous global energy minimum.  An unusual feature of the model is that it allows for ``surface hopping without the hops": the relative weights of contributing atomic states allow for chemical potential equalization-controlled switching between potential energy surface crossings, and thus, correct dissociation to neutral atoms.\cite{samuels2025}  Viewed from a machine learning perspective, the latent space patterns can serve as the starting point for constructing parsimonious potential models, with only a small number of parameters needing to be tuned in order to model a new system.\cite{valone2006electron}  
 
 The physics-based latent space 
 patterns consist of atoms dressed by quantum mechanical interactions through an ensemble representation, embedded in their (local) chemical environments, and coupled by long-range (global) electrostatic interactions. 
 The transmission of quantum mechanical information from the electronic to the atomic length scale is effected through 
 the theorems of ensemble density functional theory (DFT),\cite{kohn1986,gross1988var,gross1988DFT,perdew1982} spherical DFT,\cite{theophilou2018,nagy2018,samuels2026information} and formal, analytic results\cite{Kato1957,Steiner1963,levy1976,Tal1978} governing the short- and long-range limits of neutral, ionic, and excited state sphericalized atomic charge densities. The exclusive use of atom-centered components to define the latent space patterns avoids potential errors from molecular data, and cross-dataset inconsistencies deriving from the use of different levels of theory (exchange-correlation energy functional and basis set) in computing training set energies, densities, and forces. 
 
The ECT-EAM potential model based on the constructive latent space elements possesses several important features. The electron density provides the critical link between the electronic and atomic length scales: it carries information on the quantum mechanics of interacting \textit{electrons} into a classical interatomic potential model of interacting \textit{atoms}. As a consequence of the theorems of DFT, the nuances of quantum entanglement and electron correlation are preserved, encoded within the three-dimensional electron density scalar fields of individual atoms $\{\rho_i({\bf r})\}$, rather than a $3N_e$-dimensional, 
wavefunction $\Psi({\bf r},{\bf r}_2, \ldots {\bf r}_{N_e})$ spanning the entire $N_e$-electron system, or distributed sparsely and implicitly across energy-, density-, and force-sampled training datasets. In principle, the latent space representation can enable scaling to very large systems by virtue of utilizing atom-centered, spherically-symmetric atomic basis densities, avoiding the need to explicitly impose equivariance\cite{batzner2023} via parameterized geometric expansions with large numbers of fitting parameters. The constructive latent space approach also opens the door to the exploration of hybrid ensemble-graph neural network models for improved fidelity across a  broader array of complex molecular and materials systems,  with far fewer fitting parameters.  
 
The remainder of the paper is organized as follows. Section~\ref{sec:Background} provides backgroumd and context for the constructive latent space formulation, including an overview of the major practical and conceptual challenges faced in interatomic potential design, and a review of previous work on physics-based approaches to interatomic potential design. Section~\ref{sec:latent-space} presents the theoretical rationale for each of the components of the latent space representation, and describes how they can be unified within the design of an ensemble charge-transfer potential. The paper concludes in Section~\ref{sec:Discussion} with a discussion of next steps in assessing and applying the latent space representation, and suggests ways in which the latent space perspective can inform the design of compact, scalable, and interpretable machine-learned potentials.

\section{Background}
\label{sec:Background}
This section outlines the principal challenges of interatomic potential design, and sets the stage for discussing the latent space representation by summarizing previous physics-based approaches to empirical potentials and ML models.

\subsection{Interatomic potential design challenges}
\label{sec:DesignChallenge}
Three aspects of interatomic potential design contribute to its difficulty in practice: the combined exponential scaling of structure space and the quantum mechanical Hilbert space characterizing electron correlations; 
the challenge of interpretability arising in highly-parameterized ML models; and the question of how to accurately compress the complex quantum mechanical electron correlations information from a system's many-electron wavefunction into a more tractable and scalable density functional theory representation.

\subsubsection{Exponential scaling, or the "curse of dimensionality"}
One of the main motivations for constructing a classical atomistic potential is to enable the study of very large systems---scaling to hundreds of thousands of atoms in studies of solvated proteins, or even billions of atoms, as in the study of fracture and dislocation dynamics in mesoscale materials. A second, complementary driver is the impetus to explore greater compositional diversity in order to design technological materials with specified properties. Examples include the rational design of strong yet ductile refractory high entropy materials, high-$T_c$ superconductors, and exotic nanoscale materials. The challenge arises from the combinatoric complexity of modeling different structures, compositions, and numbers of distinct elements, particularly when there is 
off-lattice nuclear relaxation\cite{bijjala2026elastic} or other local symmetry-breaking.\cite{zunger2022bridging}.

Now imagine that one wishes to construct a ML interatomic potential by computing and fitting the energies of sampled structures comprising a total of $N_A$ atoms in 3D space. Each atom in a structure is represented by a point in $\mathbb{R}^{3N_A}$. To sample just 1\% of the overall configuration space in constructing a training dataset would require sampling some fraction $x$ of points {\it in each dimension}, such that $x^{3N_A} = .01$. For $N_A=4$ atoms, $x = 46\%$; for $N_A = 10$ atoms, $x = 86\%$; for $N_A=1000$ atoms, $x = 99.8\%$. 

\begin{figure}
\vspace*{.15in}
\includegraphics[width=0.65\linewidth]{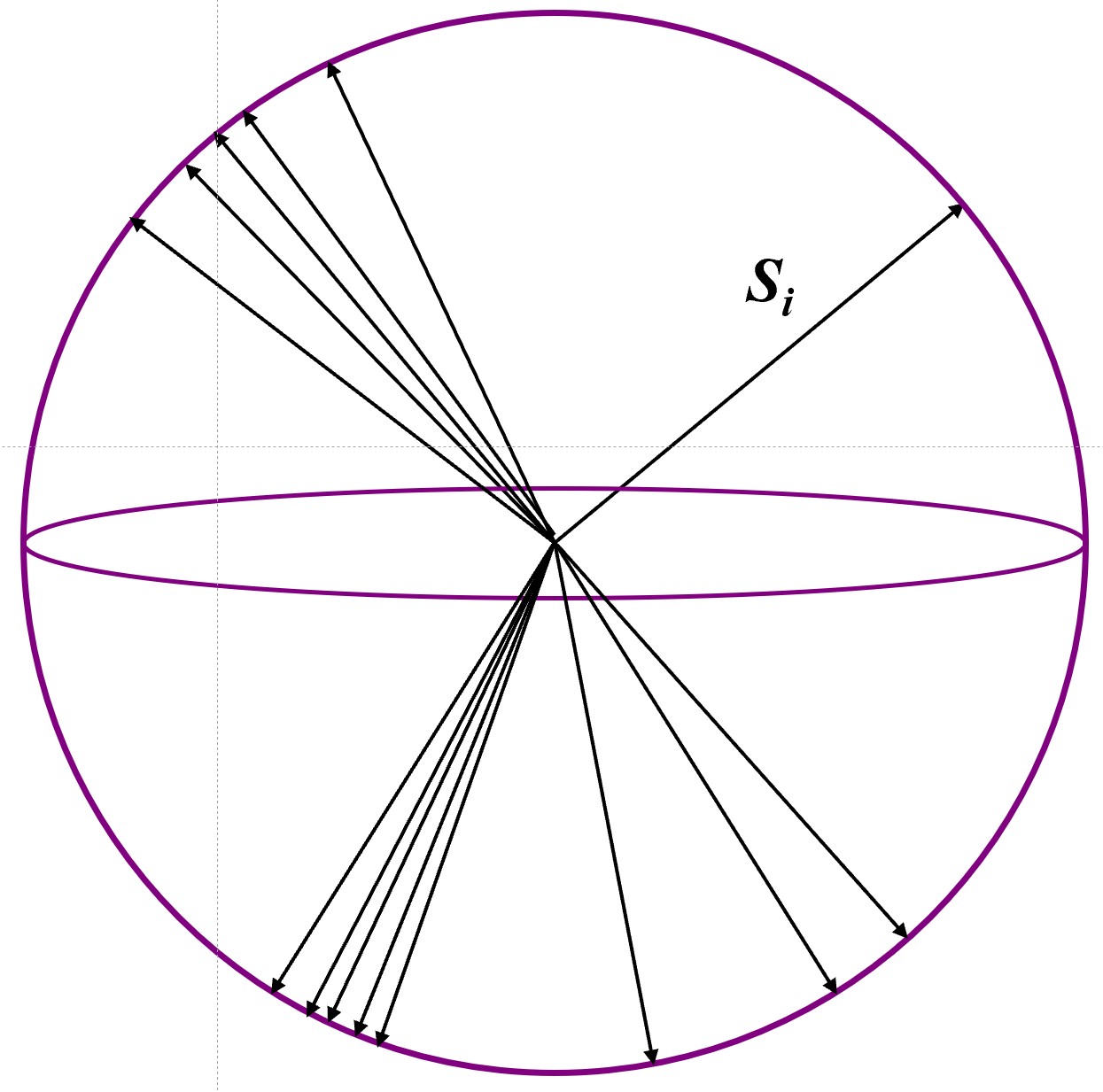}
\caption{\label{fig:CurseOfDimensionality} The ``curse of dimensionality.'' Atomic configuration vectors $S_i$ normalized to a common magnitude (sphere radius), are only sparsely distributed on the surface of the sphere in $\mathbb{R}^{3N_A}$, where $N_A$ = number of atoms. The sparse coverage of configuration space makes it difficult to discern and cluster new or rare bonding patterns.}
\end{figure}

These two examples illustrate the fundamental modeling challenge associated with interatomic potential design at the purely atomistic scale: whether due to the sheer number of atoms in the system, or the number of possible compositional proportions and atomic arrangements, the number of atomic configurations needed to achieve a specified sampling percentage of the total configurational volume grows exponentially with the number of atoms $N_A$. This phenomenon is known in the machine learning literature as {\it the curse of dimensionality},\cite{bishop1995,fn0,clarke2008properties}  and is illustrated schematically in Fig.~\ref{fig:CurseOfDimensionality}.  The sparse distribution of atomic configuration vectors makes it impossible to identify chemical patterns (``clusters") and construct a general model of the interaction potential, 
absent some pre-organized knowledge about the system.  This issue impacts both empirical potential models and machine learning approaches.

Prior to the advent of high-throughput ML techniques, empirical models were designed to describe a particular class of system and capture a limited set of expected chemical bond types and interactions using a suitable analytic form. Examples include pair interaction potentials, bond-order potentials, classical biophysical force fields, and embedded atom approaches as described in Section~\ref{sec:physics}. These models were intended to capture the essential features deriving from quantum mechanical electron-electron interactions, but in an indirect way---through effective charges, polarization and electrostatic models, and parametrized overlap integrals between wavefunctions. In ML approaches, analytical models are still needed in order to define fitting parameters, but these are typically couched in terms of mathematical constructs such as Gaussian functions, spherical harmonics, and tensor representations, in order to satisfy symmetry requirements or facilitate numerical evaluations.
The inclusion of quantum mechanical effects is implicit in the structures selected for the training dataset, and the level of theory used to compute the reference electron densities, energies, and forces to be fitted by the potential. Explicitly including QM degrees of freedom would further explode the size of the space to be modeled (empirical potentials) or sampled (ML potentials), from $3N_A$ to ${\cal{O}}(3N_A \times 3N_e)$, where $N_e$ is the maximum number of electrons associated with any single atom in the system---clearly a prohibitive prospect.  The challenge is how to allow the system itself to probe potential bonding environments dynamically, without incurring the cost of precomputed sampling that may miss relevant atomic configurations and interactions. A case in point is the debate over the appropriate quantum mechanical description of bonding in         
${\rm C}_2$, and whether it possesses unexpected quadruple bond character,\cite{shaik2012quadruple,xu2014insights,desousa2016Isthere} potentially stemming from an avoided crossing with a low-lying excited state.\cite{varandas2008extrapolation,boschen2013accurate,tobola2024reasons} 

\subsubsection{Interpretability}
As machine learning models of interatomic potentials have continued to grow in size and scope,\cite{jacobs2025practical} driven by rapid advances in graph neural network models\cite{batatia2025foundation,wood2025family} and the ability to train on increasingly large and diverse datasets\cite{curtarolo2012aflowlib,saal2013materials,zagorac2019recent,smith2020ani,horton2025accelerated,levine2025open} (see also the compilation in 
[\onlinecite{yuan2025foundation}]), the number of model fitting parameters has exploded into the millions\cite{batatia2025foundation,chorna2025comparing} and even billions\cite{wood2025family} for foundation models. In tandem, many models have begun to impose equivariance\cite{batzner2023} to ensure that models can recognize previously-learned rotationally- or translated-patterns within new data. These computational drivers have resulted in representations that have moved further away from physics- and chemistry-based design and interpretability, the original motivations for empirical\cite{tersoff1986new,brenner1990empirical,daw1983,daw1984,baskes1987,baskes1989,baskes1992modified,streitz1994,elstner1998self,zhou2004modified,senftle2016} and earlier machine learning\cite{behler2007,bartok2010gaussian,behler2011,thompson2015spectral,smith2017,wood2018extending,zhang2019embedded,behler2021four,takamoto2022} models. 

In response to the growing complexity and mathematical abstraction of machine learning models more broadly, methods such as Shapley value estimation,\cite{shapley1953value} originating in game theory, and visual attribution maps such as t-SNE\cite{van2008visualizing} have been utilized for feature selection and explainability, and to explore interpretability-accuracy tradeoffs.\cite{lundberg2017unified}  As the scale of deep learning models has grown, the field of explainable AI (XAI) has focused increasingly on methods for \textit{mechanistic interpretability}---understanding why a model makes the predictions it does.\cite{longo2024explainable,zhao2024explainability}  For MLIPs, similar efforts are underway.\cite{oviedo2022interpretable,chorna2025comparing} In light of the graph neural-network architecture of these models,\cite{duval2023hitchhiker} with atoms resident at nodes of the network, recent work has explored the possibility of  reverse-engineering \cite{longo2024explainable} the model onto properties associated with the constituent atoms.\cite{chorna2025comparing} 
These efforts are providing an opportunity to explore the convergence between atom-projected properties derived from ML models, and physics-based models designed to incorporate chemical and physical constructs {\it a priori}, within the design of the ML model itself. 

\subsubsection{The electron correlation problem}
An additional concern is the nature of the data on which interatomic potential models are trained. Wavefunction-based quantum chemistry calculations of molecular energies scale exponentially with the number of electrons $N$ in the system,\cite{helgaker2013molecular} and as $N^7$ in the 
CCSD(T) approximation,\cite{bartlett2007coupled} the current standard for quantum chemistry benchmarks and reference calculations. 
Consequently, high-throughput training datasets for molecules are typically generated using density functional theory (DFT) methods, which scale as $N^3$ and provide a good balance between efficiency and accuracy.  Most quantum mechanical calculations of materials properties  
also rely on DFT.\cite{gonze2020} The fundamental chemistry of a system therefore hinges upon the quality of the approximation to the exchange-correlation energy $E_{\rm xc}[\rho({\bf r})]$---the universal functional of the electron density whose existence is guaranteed by the Hohenberg-Kohn and Kohn-Sham theorems,\cite{hohenberg1964,kohn1965} but whose exact form is unknown. In practice, $E_{\rm xc}[\rho]$ must be approximated, and hundreds of functionals have been been developed and implemented in contemporary electronic structure codes.\cite{lehtola2018recent}

The difficult nature of the electron correlation problem stems from the two-body Coulomb repulsion operator appearing in the Hamiltonian, $1/|{\bf r} - {\bf r}'|$. These pairwise electron-electrom interactions are mediated by the many-electron wavefunction $\Psi({\bf r}, {\bf r}_2, \ldots {\bf r}_N)$, which describes the collective spatial distribution of all electrons in the system. Models of $E_{\rm xc}[\rho]$ must reconcile a description of all possible pairwise interactions between indistinguishable electrons consistent with a global $N$-body wavefunction integrating to $\rho({\bf r})$:
\begin{equation}
\rho({\bf r}) = N \int |\Psi({\bf r}, {\bf r}_2, \ldots {\bf r}_N)|^2\, d{\bf r_2}, \ldots d{\bf r}_N.
\label{eq:Born-rule}
\end{equation}
Whether viewed through the lens of the DFT electronic structure problem\cite{sun2015,medvedev2017} or that of complexity theory and quantum entanglement,\cite{schuch2009computational,ding2020concept} it is clear that the design of accurate exchange-correlation functionals presents a significant and ongoing challenge to theory.\cite{cohen2012challenges} Moreover, many of the fundamental issues are deeply intertwined.\cite{maniar2024symmetry,perdew2025scan,bartlett2019}
These include delocalization error (symmetry breaking);\cite{bryenton2023delocalization} failure to correctly dissociate molecules to neutral atoms,
\cite{perdew1982,bao2018well} a capability essential for describing reactive chemistry at the atomic scale;\cite{senftle2016reaxff} self-interaction error;\cite{perdew1981,perdew2015paradox} the ability to concurrently describe both static and dynamic correlation,\cite{sinanoglu1964many,mok1996dynamical,hollett2011two} including the nuances of the long-range van der Waals (dispersion) interaction;\cite{maggs1987electronic,atlas1988density,rapcewicz1991fluctuation,lein1999toward,vydrov2010nonlocal,grimme2016dispersion} energy tradeoffs in hybrid functionals that empirically mix Hartree-Fock and DFT correlation models;\cite{becke1993density,chai2008systematic,kumar2024critical} and limitations in computing accurate excited state energies and densities due to the use of adiabatic approximations to the exchange-correlation energy.\cite{lacombe2023non} The appropriate framing of the correlation problem can be influenced by interactions between electronic and structural (atomic) degrees of freedom---for example, energy stabilization of a material resulting from ``nesting" of local symmetry-breaking atomic motifs within a symmetric crystal structure.\cite{xiong2025symmetry} In such cases, a problem that initially appears to involve difficult-to-describe strong correlations may need to be mapped onto a conventional description of correlation in a symmetry-broken representation.\cite{zunger2022bridging,perdew2021interpretations} 

These missing pieces of the electron correlation puzzle affect even the best contemporary exchange-correlation functionals. Recent studies have highlighted inconsistencies between levels of theory\cite{karton2025good} and across databases.\cite{hegde2023quantifying}  A related issue is that different levels of theory are considered ``best" depending on the subdomain of molecular chemistry\cite{smith2017ani,smith2020ani} or materials science. \cite{Jain2013,kingsbury2022performance,horton2025accelerated,levine2025open} 
These considerations directly impact the fidelity with which interatomic interactions can be learned from large-scale DFT datasets. 

\subsection{Physics-based approaches to interatomic potential design}
\label{sec:physics}
In just a few short years, machine learning interaction potentials based on geometric graph neural networks (GNNs), and trained on large datasets of quantum mechanical calculations, have been applied to a broad array of molecular and materials simulation problems.\cite{jacobs2025practical} 
Despite remarkable progress, however, fundamental challenges in interatomic potential design remain, and these cannot be overcome merely by scaling to larger training datasets. 

In the early days of classical simulation, simple empirical models sufficed to describe the essential features of a potential intended for modeling a particular system. However, as systems have grown in chemical and structural complexity, both empirical potential models and first-generation physics-based machine learning potentials have had to evolve. This section provides a brief overview of these potentials, since they embody key aspects of the chemistry and physics of interacting atoms that can be abstracted into a pre-organized constructive latent space as proposed here. This leads to a remarkable simple picture in which the purely mathematical ML representation of an atom interacting with its environment---expressed in terms of tensors, rotation matrices, Clebsch-Gordan coefficients, Gaussian functions of radial coordinates, with associated fitting parameters---can be mapped instead onto an interpretable physics-based representation.

\subsubsection{Empirical models}
The development of empirical interatomic potentials is over a century old, dating back to the original Lennard-Jones (LJ) two-body potential,\cite{jones1924determination} expressed as competing attractive and repulsive contributions, each varying inversely with internuclear separation $R$. 
The Lennard-Jones potential model predated the publication of the Schr\"{o}dinger equation by approximately two years.\cite{schwerdtfeger2024100}  The connection with quantum mechanics, through the $1/R^6$ London dispersion interaction\cite{london1930theorie} and higher order perturbation theory  terms,\cite{london1937general} was established soon after,
leading to the 6-12 potential still in use today for modeling liquids and nanoscale metal clusters.\cite{schwerdtfeger2024100} Damped pairwise dispersion interactions\cite{tang1984improved} designed to address divergences at short $R$ arising from interpenetrating quantum mechanical charge distributions,\cite{brooks1952convergence,dalgarno1956representation,koide1976new,rosenkrantz1985damped,atlas1988density} are 
now routinely incorporated into density functional calculations in the form of post-Kohn-Sham ``DFT+D3" and related dispersion corrections.\cite{grimme2010consistent} 

Due to the profound importance of interatomic forces for understanding the states and properties of matter, empirical potentials have evolved continuously over the past century. Collectively, this work has provided insights into the broad character of interatomic forces, and the relevant parameters governing interatomic interactions.  The structure of these earlier potentials, and the physical basis for their design, are often echoed in later empirical forms. Important landmarks---by no means a complete list---include the Morse potential\cite{morse1929diatomic} and Buckingham exp-6 potential;\cite{buckingham1938classical} 
the Tersoff-Brenner bond-order 
potential\cite{tersoff1986new,brenner1990empirical}
and generalized bond-order potentials of Finnis and Sinclair\cite{finnis1984simple} and Pettifor and co-workers;\cite{pettifor1989new,pettifor2004interatomic}
valence-bond approaches;\cite{warshel1980empirical,schmitt1998multistate,morales2004new,valone2006ED}
and the ReaxFF reactive force field potential.\cite{van2001reaxff,chenoweth2008reaxff,senftle2016reaxff} Of particular note is the family of self-consistent tight-binding methods, which combine an empirical tight-binding parameterization with a density functional perturbation treatment of charge fluctuations.\cite{elstner1998self,frauenheim2002atomistic}

In parallel with these developments, empirical force field models specifically tailored for macromolecular simulation have been under continuous development since the pioneering work of the Karplus group in the 1980s,\cite{brooks1983charmm} due to their importance in biophysical modeling and drug discovery. These models assume a strictly-defined amino acid sequence in which covalent peptide bonds between amino acids impose an effective carbon atom backbone manifold constraint. The interatomic forces are described in terms of harmonic bonds and angles, dihedral and torsional angle terms, and classical Coulomb and Lennard-Jones long-range (``non-bonded") interactions.  Prominent examples include the CHARMM\cite{brooks2009,hwang2024charmm} and Amber\cite{ponder2003force,wang2025evolution} force fields.  

The most important class of empirical models for purposes of the present discussion are those based on the {\it embedded atom method} (EAM), originally proposed by Daw and Baskes.\cite{daw1983,daw1984}  Motivated by effective medium theory,\cite{stott1980,norskov1980,norskov1982} EAM partitions the many-body potential into two components: an atom-centered embedding term, unique to each atom type in the system, which encapsulates effects due to the local, quantum mechanical bonding environment of each atom; and an electrostatic interaction term, which captures long-range classical electronic interactions. The mathematical formulation of the original EAM is given in Section~\ref{sec:latent-space}, as part of the discussion of the ensemble charge-transfer EAM potential.\cite{muralidharan2007ED,Atlas2021}

The original EAM proved remarkably successful in describing fcc metallic systems.\cite{foiles1986,mishin2001}  Over the years, numerous empirical modifications were proposed to address covalently-bonded materials (the modified embedded atom method, MEAM),\cite{baskes1987,baskes1989,baskes1992modified} charge transfer in metal oxides,\cite{streitz1994,zhou2004modified} incorporate van der Waals interaction corrections,\cite{baskes1999many} and include bond-order effects (MEAM-BO).\cite{mun2017interatomic}  These successive improvements can be seen as mirroring the development of other empirical potentials described above, albeit within a single empirical framework.  For example, one of the major innovations of MEAM was to incorporate an environment-dependent expansion of the electron density centered at each atom, with angular dependence and tunable parameters, replacing the exponential densities of the original EAM.\cite{foiles1986}  The idea of requiring that an atom's electron density adjust as a function of its instantaneous atomic environment has subsequently been seen as an essential ingredient of MLIPs, as well as motivating the ``energy-density duality'' design of the ensemble charge-transfer EAM.\cite{muralidharan2007ED,Atlas2021}  

\subsubsection{Physics-based machine learning models}
\label{sec:ML-models}
\begin{figure*}
\centering
\includegraphics[width=0.47\linewidth]{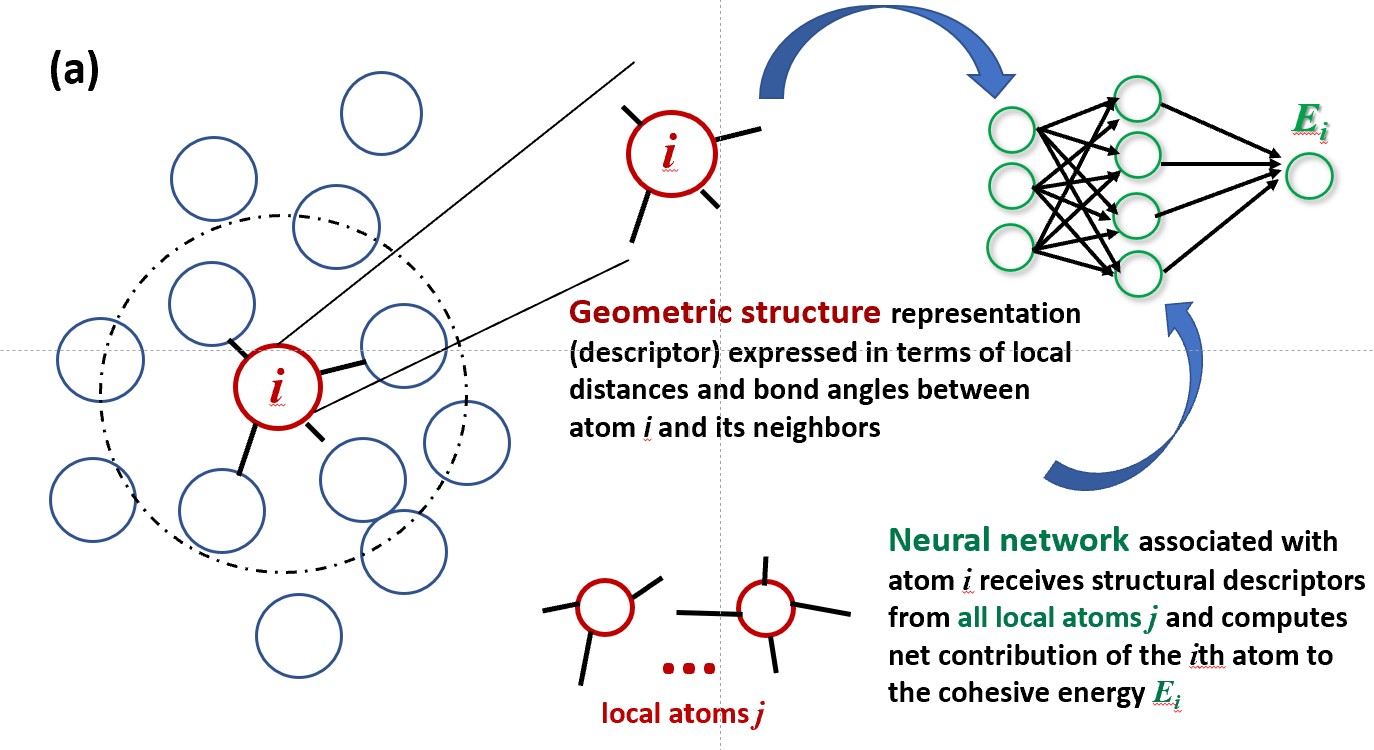} \hfill
\includegraphics[width=0.52\linewidth]{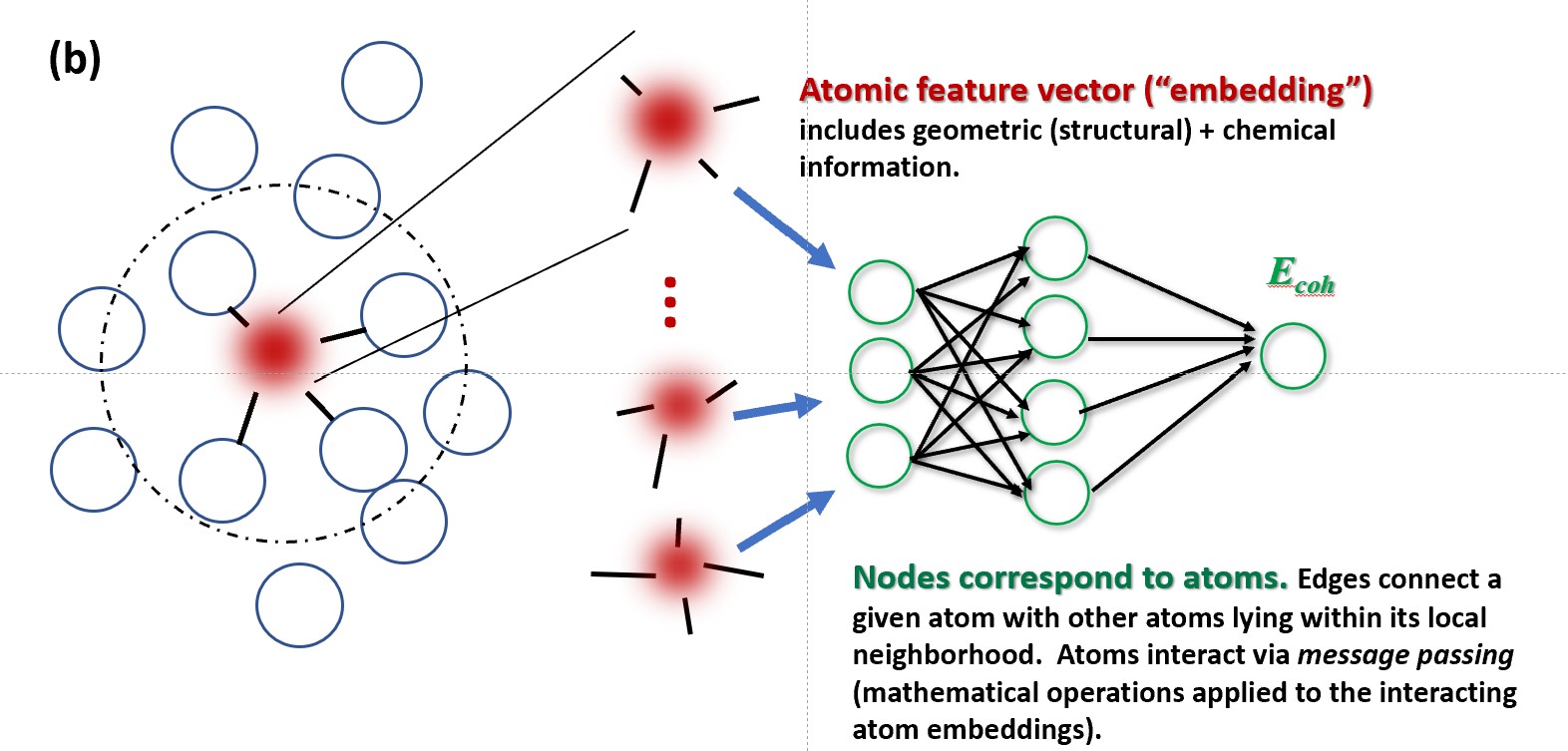}
\includegraphics[width=0.68\linewidth]{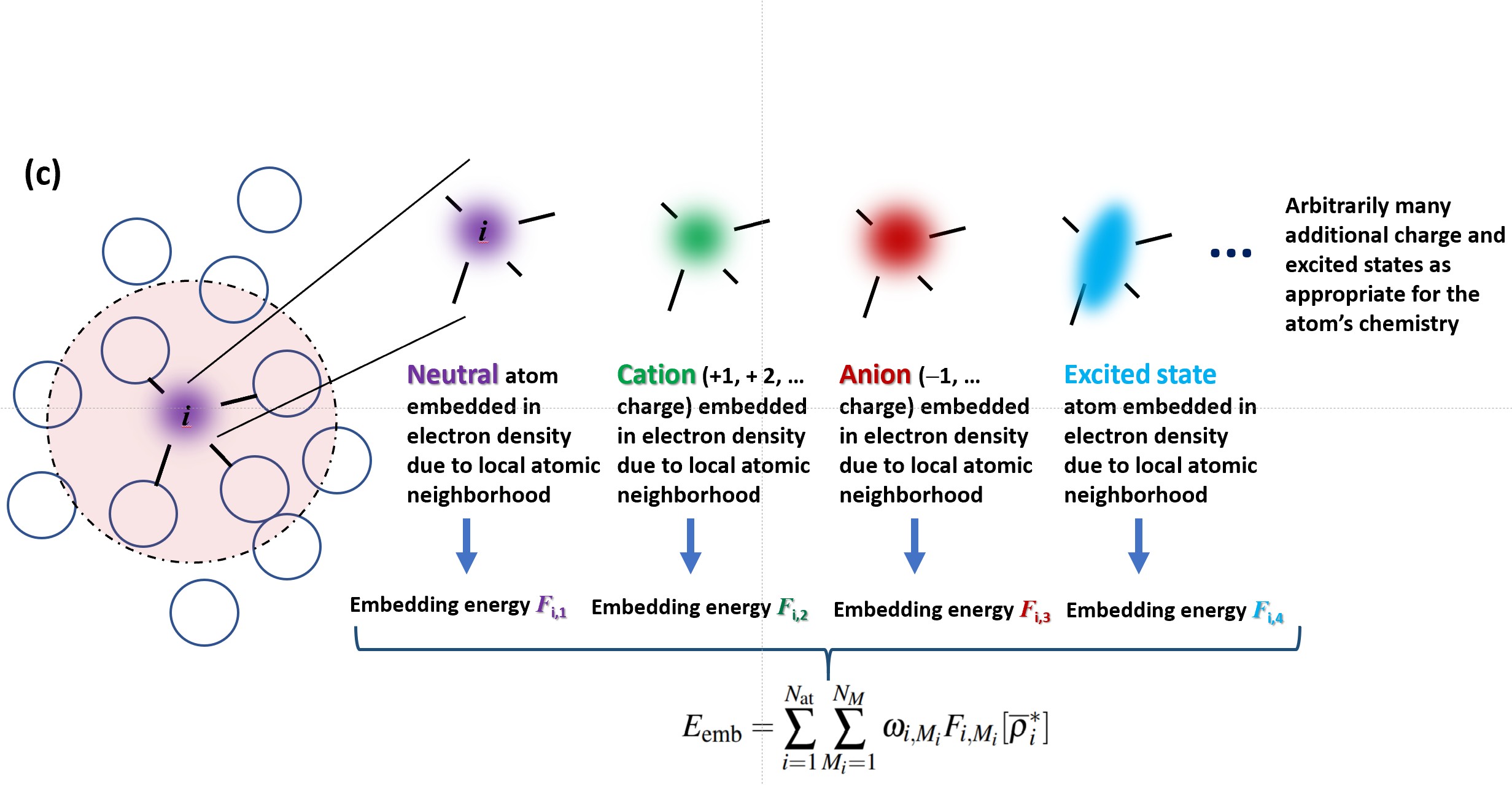}
\caption{\label{fig:Potentials} (a) Schematic representation of Behler-Parrinello\cite{behler2007} and ANI-type MLIPs.\cite{smith2017} Geometric information about the local environment of an atom is encoded in a descriptor comprised of local symmetry functions. A clear separation exists between geometry (the atomic descriptor, red circles) and the physico-chemical interactions (the neural network, in green). (b) Generic graph neural network (GNN) architecture. Atoms (fuzzy red disks) can encode both geometry and chemical information (atomic number, electronegativity, etc.)~within a descriptor, termed an \textit{embedding}.  The atoms corespond to nodes of a neural network, and interactions between the atoms are encoded as ``messages" (mathematical operations on the atom embeddings), indicated by edges in the GNN.  In this architecture, geometry and chemistry are effectively convolved.  (c) Constructive latent space representation (present work).  In the ensemble charge-transfer embedded atom (ECT-EAM) interatomic potential based on this representation,\cite{muralidharan2007ED,Atlas2021} each atom $i$ contributes an embedding energy consisting of the weighted sum of state embedding energies $F_{i,M_i}$, where $M_i$ is an index ranging over all contributing isolated atom states: ground state (purple), positive (green) and negative (red) ions, and excited states (aqua). The positive ion is indicated with slightly contracted disk, and the negative ion with a more diffuse representation. The excited state is indicated with a distorted density representation, as a visual suggestion of charge polarization. The embedding energy function for each atomic state is computed as described in the text.  The net embedding energy contribution from atom $i$ is the weighted sum of the state-specific embedding functions.  The statistical weights $\omega_{i,M_i}$ appearing in the ensemble are required to sum to 1, and provide a measure of the importance of the $M_i$th state to the overall embedding energy contribution from atom $i$. The sum over all atoms in the system gives the total embedding energy $E_{\rm emb}.$  The additional electronstatic contribution to the total cohesive energy (interatomic potential) is computed as described in the text; see Eq.~(\ref{eq:ensFF}).}
\end{figure*}

The current era of machine learning models for interatomic potentials began with the pioneering work of Behler and Parinello in 2007.\cite{behler2007} The total cohesive energy (interaction potential) was computed as the sum of individual atomic components, $E_i$ (see Fig.~\ref{fig:Potentials}(a)):
\begin{equation}
    E = \sum_{i=1}^{N_{\rm A}} E_i.
    \label{eq:atomic-sum}
\end{equation}
A similar design was subsequently implemented by Isayev, Roitberg, and co-workers.\cite{smith2017} In this architecture, the chemical environment of the $i$th atom is characterized in two steps. First, the local \textit{geometry} is expressed in terms of a set of atom-specific angular and radially-dependent symmetry functions ${\cal G}_i^{\mu}$ centered at $i$. These atom-centered symmetry functions (ACSFs)\cite{behler2011} are controlled by a radial cutoff function, yielding a local geometric representation overall.  In the second step, the local geometry information from the $\{{\cal G}_i^{\mu}\}$ is input to an atom-specific neural network $S_i$, which computes the $i$th atom's energy contribution $E_i$. A key point, as illustrated in Fig.~\ref{fig:Potentials}(a), is that this design distinguishes local geometry from local chemistry, facilitating the imposition of geometric invariance constraints. 

Subsequent generations of neural network-based interaction potentials have built upon this basic idea, with proposed improvements to the geometric descriptors characterizing local structure, as well as enhancements aimed at improving total energy predictions.  The recent introduction of tensor-based models has focused on improving geometric descriptions to enable more accurate and complete representations of atomic and system-wide symmetries, including equivariance. These models generalize the ACSFs from user-crafted functional forms\cite{behler2007,smith2017} to more abstract and complex geometric representations variously expressed in terms of spherical harmonic expansions, products of spherical harmonics coupled by Clebsch-Gordan coefficients, and invariant polynomials, combined with a local radial basis, and with the advantage of facilitating scalability with the number of atoms and different element types. A compilation of early descriptors (through 2019) can be found in [\onlinecite{behler2021four}]. Prominent examples include the spectral neighbor analysis potential (SNAP) method, which uses bispectrum components projected onto hyperspherical harmonics to describe local angular dependence;\cite{bartok2013,thompson2015spectral,wood2018extending} the invariant polynomials of moment tensor potentials,\cite{shapeev2016moment} and the atomic cluster expansion (ACE) approach\cite{drautz2019atomic,dusson2022atomic} used in the MACE equivariant message-passing neural network architecture\cite{batatia2022mace,batatia2025foundation} (a schematic illustration of the graph neural network architecture for interatomic potentials can be found in Fig.~\ref{fig:Potentials}(b)). Importantly, equivalences between seemingly different atomic environment representations (embeddings) can be established in terms of the ACE representation,\cite{drautz2019atomic,batatia2022mace,nigam2022unified,darby2023tensor} through appropriate transformations and limits.\cite{drautz2019atomic,musil2021physics}

Comparisons of subsequent generations of MLIPs utilizing these highly-developed atomic embeddings can be found in recent research papers and reviews.\cite{drautz2019atomic,zuo2020performance,behler2021four,unke2021machine,duval2023hitchhiker,yuan2025foundation}  Before turning to a description of the constructive latent space representation in Section~\ref{sec:latent-space} (illustrated in Fig.~\ref{fig:Potentials}(c)), it will be helpful to comment on how electrostatic effects are incorporated into various MLIPs, and briefly discuss a number of atomic embeddings that emphasize encoding of physical and chemical information at a deeper level than simple scalar descriptors such as atomic number and electronegativity, as often used in geometric structure-focused models. Both of these issues will be revisited below.

The MLIPs that we have highlighted thus far make an assumption of ``nearsightedness.'' This term was 
originated by Nobel laureate Walter Kohn\cite{kohn1996density,prodan2005nearsightedness} to aid in understanding local chemical bonding and chemical transferability\cite{fias2017chemical} in many-electron systems, ``in the absence of long-range ionic interactions."\cite{prodan2005nearsightedness}  However, if a MLIP is to be broadly applicable, the potential must go beyond the assumption of short-range interactions, and address the electrostatic effects that are ubiquitous in biophysical systems and also arise in many materials science applications, such as reactions at surfaces.  In addition to constructing MLIPs based on atom-centric embeddings as illustrated in Fig.~\ref{fig:Potentials}, several groups have also worked on incorporating charge-transfer effects. These will have a direct analog in the ensemble electrostatic contribution in the latent space model to be discussed below.  One recent example of a MLIP augmented to address electrostatic effects is the 4th generation neural network model of Behler,\cite{behler2021four} in which two separate neural nets---one describing short-range energy contributions including local atomic charges, and the other describing long-range electrostatic interactions---are coupled through a charge equilibration procedure utilizing environment-dependent atomic electronegativities.  This design was based on the charge equilibration neural network technique (CENT) of Ghasemi {\it et al.}, which implemented a local harmonic fluctuating charge model about each atom, in the spirit of the Streitz and Mintmire's ES+ extension to the EAM.\cite{streitz1994} Recent work by 
Shaidu {\it et al.}~retained the fluctuating charge model and dependence on local electronegativities, but in a simplified, single neural network model.\cite{shaidu2024incorporating}  In the PhysNet model, Unke and Meuwly augmented the predicted short-range energies from a message-passing graph neural network model with electrostatic interactions computed between between PhysNet-predicted charges.\cite{unke2019physnet} A similar approach was used in AIMNet2, where the total potential was written as the sum of local, electrostatic, and dispersion terms.\cite{anstine2025aimnet2} AIMNet2 and its predecessor\cite{zubatyuk2019} are also notable for introducing an ``atom-in-molecule inspired" numerical descriptor designed to capture chemical neighborhood information in the form of learned features associated with each atom, in addition to the atom's local \textit{geometric }information.  In AIMNet2, the chemical descriptor includes effective point charges, which are used to compute the electrostatic potential contribution. A review of the treatment of electrostatic interactions using charges computed from atomic descriptors can be found in
[\onlinecite{anstine2023machine}].

Several groups have developed embedded-atom-inspired neural network representations of the interatomic potential. Zhang \textit{et al.}~proposed EANN, in which the embedding density at each atom is modeled using Gaussian-type orbitals, and a neural network is designed to replace the embedding functions.\cite{zhang2019embedded}  Inspired by this work, Takamoto and coworkers developed the tensor embedded atom
network (TeaNet), aimed at capturing the physics of Baskes' modified embedded atom method (MEAM)\cite{baskes1987,baskes1989} by representing MEAM's bond-order and angular dependence in a graph convolutional network architecture.

An important development was Drautz's introduction of the atomic cluster expansion (ACE),\cite{drautz2019atomic} subsequently adopted for use by the MACE series of MLIP models.\cite{batatia2022mace,batatia2025foundation}  Drautz showed that the ACE description of the local atomic environment was sufficiently general as to encompass Tersoff\cite{tersoff1986new} and Finnis-Sinclair\cite{finnis1984simple} bond-order potentials; MEAM's angle-dependent expansion of the electron density;\cite{baskes1987,baskes1989,baskes1992modified} the bispectrum dependence of the SNAP potential;\cite{thompson2015spectral} and Shapeev's moment tensor potentials with angle-dependent densities corrected to higher order beyond MEAM.\cite{shapeev2016moment}  Drautz also proposed an analytic ``cluster functional" form for generalizing the Finnis-Sinclair $\sqrt{\rho}$ functional form that plays a role analogous to the embedding functions of the EAM.  It is interesting to note in this context that several decades earlier, Moriarty\cite{moriarty1988density} had proposed and extensively developed cluster expansion (``multi-ion'') interatomic potentials derived from DFT generalized pseudopotential theory (GPT), to account for angular forces in the bcc transition metals, and proposed a model analytic form (MGPT)\cite{moriarty1990analytic} to facilitate use of the cluster expansion potential in molecular dynamics simulations. 

This very brief and necessarily incomplete overview of MLIPs has highlighted the connections between embedded atom-type methods, bond order empirical potentials, and DFT-based cluster expansions leading up to the design of contemporary graph neural network potentials, with physics-based models implemented at the (atomic) nodes. In Section~\ref{sec:latent-space} we will show how ensemble density functional theory makes it possible to unite local atomic models within a rigorously formulated interatomic potential, the ECT-EAM, couched in terms of constructive latent space DFT components. This opens the prospect of merging the DFT-based latent space representation with a graph neural network to construct a parsimonious but generalizable MLIP.

We conclude this section by mentioning a number of  novel chemistry-based representations of the local atomic environment that have been proposed in the literature. Huang and von Lilienfeld introduced the concept of an extended Periodic Table expressed in terms of chemical fragments of increasing size (``amons") corresponding to common chemical environments, as the basis for their \textit{amon-based machine learning model} (AML).\cite{huang2020} Billinge\cite{billinge2024materials} explored the idea of utilizing the 1D atomic pair distribution function (PDF)---a histogram of the ordered interatomic distances for a set of atoms in a material---to replace the standard set of atomic 3D coordinates, and define the atoms' ``genetic code."  The reverse problem---converting a complete list of interatomic distances to a set of 3D coordinates that define an embedding of atoms in 3D space---is known as the \textit{unassigned distance geometry problem} (UDGP).\cite{duxbury2016unassigned}  The UDGP provides an intriguing link between the idea of using the PDF to define the local geometry of an atom, and the sphericalized atomic state densities that are part of the DFT constructive latent space representation (see Section~\ref{sec:latent-space}.) The connection derives from a theorem of distance geometry,\cite{havel1983theory} recently used to extend\cite{samuels2026information} spherical DFT\cite{theophilou2018,nagy2018} to cases where the centers 
$\{{\bf R}_i\}$ of the set of spherical DFT atomic densities are unknown. The relationship can be expressed as:
\[
    {\rm \ PDF\ genetic\ code}\ \Longrightarrow  \{{\bf R}_i\}\ \Longleftarrow {\rm spherical\ atomic\ densities}  
\]
In both representations, the set of all interatomic distances suffices to encode the local atomic geometry information, including implicit angular dependencies.

\section{Constructive latent space representation}
\label{sec:latent-space}

The concept of an encoder-decoder machine learning architecture (subsequently dubbed an \textit{autoencoder}\cite{hinton2006}) was described by Ackley, Hinton, and Sejnowski in 1985.\cite{ackley1985learning}  Quoting from their paper:
\begin{quote}
``... there must be some way of choosing internal representations which allow the preexisting hardware connections to be used efficiently for encoding the constraints in the domain being searched... We describe some simple examples in which the learning algorithm creates {\it internal represenations} [emphasis ours] that are the most efficient way of using the preexisting connections."\cite{ackley1985learning}
\end{quote}
Following the demonstration of effective learning methods for multilayer neural networks based on backpropagation,\cite{rumelhart1986learning} Hinton and Salakhutdinov demonstrated 
\begin{quote}
``an adaptive, multilayer `encoder' network to transform... high-dimensional data into a low-dimensional code and a similar `decoder' network to recover the data from the code... The whole system is called an `autoencoder'..."\cite{hinton2006}  
\end{quote}
The learned encoder is represented by successive layers in the neural network, and generates the lower-dimensional {\it latent space} encoding of the data.\cite{bank2023autoencoders} The resulting compressed representation can be used to reconstitute the original data, and recognize new data not yet seen.  While there is some loss of detail in the reconstruction, the key point is that the essential aspects of the data (the ``eight-ness" of the number 8, for example),\cite{hinton2006} are learned and abstracted into the latent space representation.  

\begin{figure*}
\includegraphics[width=1.0\linewidth]{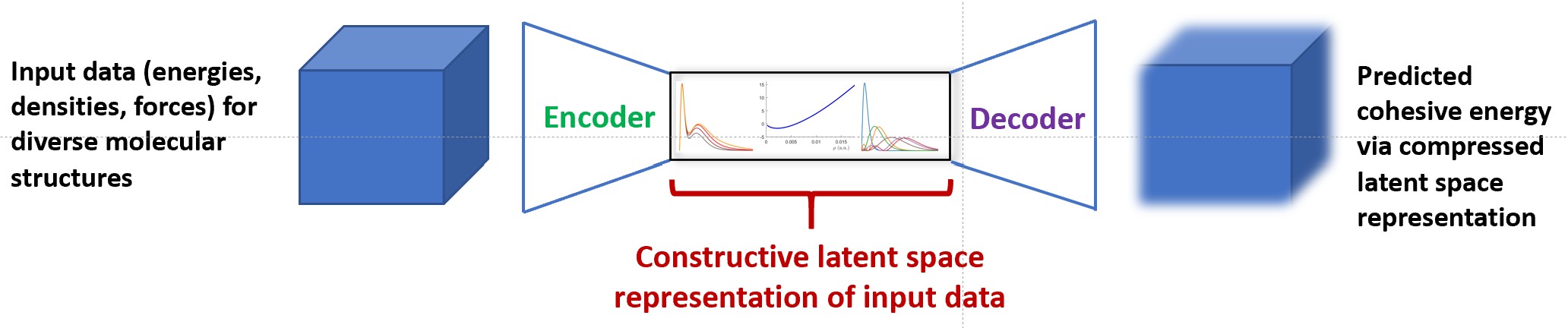}
\caption{\label{fig:latentspace} Schematic illustration of an encoder/decoder architecture implementing a physics-based, constructive latent space representation of a molecular interatomic potential. The box at left corresponds to a database of energies. densities, and forces computed for a set of exemplar molecular configurations. The constructive latent space representation is specified {\it a priori}, in terms of density functional constructs---ensembles of atomic radial basis functions (RBFs) and embedded atom functions $F_i$---satisfying known physical constraints. The fuzzified box at right corresponds to the same input configurational dataset, but with molecular energies, densities, and forces computed from the lower-dimensional latent space representation. Since the latent space structures compress input molecular information onto a lower-dimensional representation (manifold), the decoded version of the data will contain errors (indicated by fuzzy edges).  A high-quality latent space representation will nevertheless preserve essential molecular characteristics such as bond orders, effective charges, and dissociation behavior (indicated here by the retention of the original cube geometry.)  By contrast, in a conventional autoencoder architecture,\cite{hinton2006} the latent space representation is learned through iterative optimization of neural network parameters to numerically reproduce digitized data, and the identification and interpretation of any emergent structure contained within the learned latent space must be deduced via downstream analysis techniques. After [\onlinecite{bank2023autoencoders}].}
\end{figure*}

The thesis of the present work is that by encoding \textit{chemical and physical understanding} of the intrinsic quantum mechanical structure of an interatomic potential in a formally-justified, {\it constructive} latent space representation---expressed in terms of ensemble density functional theory, density constraints, theoretically-grounded exchange-correlation functionals, and the density functional origins of the embedded atom method---one can design more accurate and scalable interatomic potentials applicable to both molecules and materials. A schematic of the constructive latent space architecture is shown in Fig.~\ref{fig:latentspace}. 

The constructive latent space approach provides a new route for addressing the design challenges summarized in Section~\ref{sec:DesignChallenge}: the curse of dimensionality (sampling of large configurational datasets is not needed); interpretability (the constructive latent space is interpretable by design); and the electron correlation problem (quantum correlations are encoded within precomputed, sphericalized atomic density basis functions, the density-dependent embedding functions, and the ensemble DFT representation.) In a dynamical simulation, ``nearsighted" local contributions to the potential are balanced against long-range, global electrostatic contributions through self-consistent chemical potential equalization across the entire system, causing the ensemble weights to adjust to favor different combinations of atomic ground, excited, and ionic states.\cite{Atlas2021}

\subsection{The ensemble charge-transfer embedded atom method}
The ensemble charge-transfer embedded-atom method (ECT-EAM)\cite{muralidharan2007ED,Atlas2021} provides the starting point for discussing the constructive latent space representation. It is a classical interaction potential designed to extend the original EAM to describe charge polarization and charge transfer in interacting atomic systems.  Unlike the EAM and most of its successors, it is  not limited to periodic materials; its strictly atomistic formulation ensures that it is applicable to molecules as well. 

The theoretical foundations of the ECT-EAM interatomic potential have been discussed in detail in Ref.~[\onlinecite{Atlas2021}], so the model will be summarized briefly here. The emphasis will be on providing the theoretical rationale for each of the components contributing to the final energy expression, in order to highlight how the electron density---as a fundamental \textit{latent space variable} of the theory---establishes a principled link between the electronic (quamtum mechanical, correlated electron interactions) and classical (atomistic) length scales. The ECT-EAM does not require a machine learning or a graph neural network structure for its implementation, although there may be opportunities to explore future machine learning refinements as discussed later in Section~\ref{sec:Discussion}. It is comprised solely of interpretable components constructed from sphericalized atomic basis densities that can be computed to high precision from standard quantum codes, and density functional theory energy expressions based on standard kinetic energy and exchange-correlation functionals. 

The development of the ECT-EAM was motivated by the elegant work of Daw,\cite{daw1989} who showed that under certain assumptions, the EAM could be formally derived from density functional theory.  In the same work, Daw tested the DFT-EAM connection by constructing an EAM potential for Ni, starting from Hohenberg-Kohn DFT implemented in the local density approximation. (Of note, Daw's (non-interacting) kinetic energy functional also included a scaled von Weisz\"{a}cker gradient term, akin to functionals used in orbital-free DFT subsystem and embedding models; see Ref.~[\onlinecite{mi2023orbital}]).

The original EAM expression\cite{daw1983,daw1984} is the sum of individual atomic embedding contributions $F_i$, and a term consisting of pairwise electrostatic interactions between nuclei:
\begin{equation}
E_{{\rm coh}} = \sum_{i=1}^{N_{\rm A}} F_i(\bar \rho_i(\textbf{R}_i)) + \frac{1}{2} \sum_{i\ne j} \phi_{ij}(R_{ij}). \label{eq:EAM}
\end{equation}
$F_i$ is an element-dependent embedding function of the effective local background electron density $\overline{\rho}_i({\bf R}_i)$ at atomic site $i$, and $\phi_{ij}$ is the effective electrostatic pair potential between atoms $i$ and $j$.  In the original EAM,  model electron densities were associated with each nucleus, and parameterized along with the other functions in the model.  The embedding density $\overline{\rho}_i({\bf R}_i)$ 
was then approximated as the sum of the tails of the $n_i$ nearest-neighbor atom electron densities at site $i$:
\begin{equation}
\overline{\rho}_i({\bf R}_i) \simeq \sum_{\stackrel{j=1}{j\ne
i}}^{n_i}{\rho_j^a({\bf R}_{ij})}. \label{rhobar}
\end{equation}
where $\rho_j^a$ corresponds to the isolated atom electron density of neighbor $j$. 

A well-known limitation of the original EAM was the fixed form of the electron densities appearing in the model. Daw considered this issue from a formal perspective in his derivation of the EAM from DFT,\cite{daw1989} but the treatment was perturbative and did not address charge transfer. An early effort to generalize the EAM to address charge transfer, ES+, assumed a local charge fluctuation model and used different parameterizations for the charge density in the embedding and electrostatic terms.\cite{streitz1994} 

The foundations of the EAM in DFT formalized by Daw, coupled with the Fragment Hamiltonian (FH) model of Valone and co-workers,\cite{valone2014,valone2015communication} suggested that a formal generalization of the EAM might be possible via an ensemble DFT approach.  The resulting expression for the ECT-EAM\cite{muralidharan2007ED,Atlas2021} combines ensemble DFT for excited states\cite{kohn1986,gross1988var,gross1988DFT} and ensemble DFT for charge states:\cite{perdew1982}        
\begin{equation}
E = \sum_{i=1}^{N_{\rm A}} \bigg[ \sum_{M_i=1}^{N_M} \omega_{i,M_i} F_{i,M_i}
[\overline{\rho}_i^*] + \frac{1}{2}{\sum_{j\ne i}}
\sum_{M_i=1}^{N_M} \sum_{P_j=1}^{N_P} \omega_{i,M_i} \omega_{j,P_j}\Phi_{ij,M_iP_j}\bigg].
\label{eq:ensFF}
\end{equation}
The background embedding electron density $\overline{\rho}_i^*$ at the location ${\bf R}_i$ of the $i$th atomic nucleus is given by:
\begin{equation}
\overline{\rho}_i^*({\bf R}_i) \simeq \int d{\bf r}\,  \tilde{\omega}({\bf r})\, \sum_{\stackrel{j=1}{j\ne
i}}^{n_i}{\rho_j^*({\bf r}-{\bf R}_j)},
    \label{eq:AIM-dens}
\end{equation}
where $\tilde{\omega}({\bf r})$ is a weighting function that averages over the contributions of atom-in-molecule tails in a neighborhood of atom $i$; see Fig.~\ref{fig:Potentials}(c). (The definition of the atom-in-molecule representation in ECT-EAM is given in Section~\ref{sec:latent-embedding} below). In the original EAM, it is assumed that $\tilde{\omega}({\bf r}) = \delta({\bf r} - {\bf R}_i)$.\cite{daw1989}  $M_i$ combines the atomic excited states (corresponding to charge polarization) and ionic (charge) states in a single index looping over all $N_M$ charge and excited ensemble states. A second combined index $P_j$ loops over the $N_P$ charge and excited ensemble states of interacting atom $j$.
The decomposition of the total electron density of the system in terms of atom-in-molecule densities $\rho_i^*({\bf r})$ is given by:
\begin{equation}
\rho({\bf r}) =\sum_{i=1}^{N_{A}} \rho_i^*({\bf r}).
\label{eq:DD}
\end{equation}
$F_{i,M_i}$ is the embedding function for the $M_i$th ensemble state of atom $i$; $\Phi_{{ij},{M_iP_j}}$ is the classical Coulomb interaction between the atom $i$ in state $M_i$ and atom $j$ in state $P_j$.

\subsection{\label{sec:latent-embedding} The DFT-based latent space embedding}
\noindent
Table~\ref{tab:latent-space} summarizes the key latent space components of the ECT-EAM interatomic potential form, and the rationale for characterizing them as foundational in coupling the quantum and atomistic length scales. The remainder of this section provides additional details on the representation, with notation following that of Ref.~[\onlinecite{Atlas2021}].
\begin{table*}[htbp!]
\caption{\label{tab:latent-space} Latent space components of the ensemble charge-transfer embedded atom method. Notation as described in the main text.}
\begin{ruledtabular}
    \begin{tabular}{l|l}
    {\bf Latent variable} & {\bf Rationale} \\ \hline
    1.~System electron density $\rho({\bf r})$ expressed in terms of ensemble & $-$ Born statistical 
    interpretation of QM \\
    atom-in-molecule densities $\rho_i^*({\bf r})$. & $-$ Density functional theory (DFT) \\
    & $-$ ``Disentanglement” of molecular $\rho({\bf r})$ for chemical interpretation and dyna-\\
    & \ \ \ \ \ mical charge assignment \\ \hline
    2.~Atom-in-molecule densities $\rho_i^*({\bf r})$ expressed in terms of an & $-$ Ensemble excited state and charge state DFT \\
    ensemble of sphericalized atomic state basis functions $\rho_{ijk}({\bf r})$ & $-$ Family of all possible DFT problems for a given $Z$\\
    with short- and long-range constraints. & $-$ Atomic density ``fingerprint" for given $Z$ is computed once only \\
    & $-$  Spherical DFT \\ \hline
    3.~Atomic-state-dependent energy embedding functions. & 
      $-$ Embedding concordances for ground, ionic, and excited \\ 
    & \ \ \ \ \ states of atoms \\ \hline
    4.~Electrostatic interactions, computed via the $\{\rho_i^*({\bf r})\}$. & $-$ Classical Coulombic interaction between atoms $i$ and $j$ with \{$\rho_i^*({\bf r}),Z_i$\} \\ 
    \end{tabular}
\end{ruledtabular}
\end{table*}

\vspace*{.05in}
\noindent
1. {\it System electron density $\rho({\bf r})$ expressed in terms of the ensemble atom-in-molecule densities $\rho_i^*({\bf r})$.} \hfill \\
\noindent
The Born rule of quantum mechanics\cite{born1955statistical} relates the square of the many-electron wavefunction to the total electron density of the system, as given in Eq.~(\ref{eq:Born-rule}). This provides the initial compression of information necessary for describing system-wide quantum mechanics efficiently in the ECT-EAM interatomic potential, through the relationship between $\rho({\bf r})$ and the total energy of the system given by the Hohenberg-Kohn\cite{hohenberg1964} and Kohn-Sham\cite{kohn1965} theorems of DFT.  

The next step is to use the concept of the {\it atom-in-molecule} to perform a decomposition of the electron density into atomic-like components $\rho_i^*({\bf r})$:
\begin{equation}
  \rho({\bf r}) = \sum_{i=1}^{N_{A}} \rho_i^*({\bf r}),
\label{rhoMap}
\end{equation}
where $N_{A}$ is the number of atoms in the system, and * is used to denote  atom-in-molecule quantities. The atom-in-molecule decomposition is clearly not unique, and one test of whether a given decomposition is chemically-reasonable is to evaluate the derived effective charge $q_i$ for the $i$th atom-in-molecule:
\begin{equation}
    q_i = Z_i - \int \rho_i^*({\bf r})\, d{\bf r}.
\end{equation}
This allows comparison with other decompositions such as the Bader topological atom-in-molecule approach,\cite{bader1985,bader1990} as well as testing for correct molecular dissociation in the limit of large separation between fragments.\cite{valone2004,amokwao2020,samuels2025}

\vspace*{.05in}
\noindent
2. {\it Atom-in-molecule densities $\rho_i^*({\bf r})$ expressed in terms of an ensemble of sphericalized atomic state basis functions $\rho_{ijk}({\bf r})$ satisfying short- and long-range constraints.}  \hfill \\
\noindent
In order to maintain the formally-required consistency between energies and densities in excited state ensemble DFT\cite{kohn1986,gross1988var,gross1988DFT} and charge state ensemble DFT,\cite{perdew1982} the atom-in-molecule densities $\rho_i^*({\bf r})$ are expressed as nested ensembles:
\begin{equation}
\rho_i^*({\bf r}) = \sum_{j= -\infty}^{Z_i - 1} \alpha_{ij} \varrho_{ij}({\bf r}),
\label{eq:AIM-expans}
\end{equation}
with weights $\alpha_{ij} \ge 0$ $\forall$ $i$,$j$.
The excited state ensemble density $\varrho_{ij}({\bf r})$ for atom $i$ in charge state $j$ is defined as:
\begin{equation}
\varrho_{ij}({\bf r}) \equiv \sum_{k=0}^{\infty} \beta_{ijk} \rho_{ijk}({\bf r}),
\label{eq:excit-expans}
\end{equation}
with $\beta_{ijk}$ $\ge 0$ $\forall$ $i$, $j$, $k$.  The \textit{basis density} $\rho_{ijk}({\bf r})$ is the density of the $k$th eigenstate of the $j$th ion of atom $i$ and number of electrons $N_{ij} = Z_i + j$. $j$ labels the charge state, and $k$, the excitation (polarization) state of the atom. 
$N_i \equiv N_{i0} = Z_i$ is the number of electrons in the neutral atom.  The weights appearing in the charge state ensemble Eq.~(\ref{eq:AIM-expans}) and excited state ensemble Eq.~(\ref{eq:excit-expans}) satisfy separate sum rules: $\sum_{j= -\infty}^{Z_i - 1} \alpha_{ij} = 1$ for each atom $i$, and $\sum_{k=0}^{\infty} \beta_{ijk} = 1$ for each atom $i$ in ionic state $j$. 
Combining Eq.~(\ref{eq:AIM-expans}) and Eq.~(\ref{eq:excit-expans}) gives:
\begin{equation}
\rho_i^*({\bf r}) = \sum_{jk} w_{ijk} \rho_{ijk}({\bf r}),
\label{eq:w-wts}
\end{equation}
where $w_{ijk} \equiv \alpha_{ij} \beta_{ijk}$. In light of the $\alpha_{ij}$ and $\beta_{ijk}$ sum rules, the $w_{ijk}$ satisfy the sum rule $\sum_{jk} w_{ijk} =1$ for each $i$.  Although not obvious from the compressed index notation of Eq.~(\ref{eq:ensFF}), an important feature of the ECT-EAM model is the constraint that the ensemble weights of densities and energies must correspond exactly, to ensure internal consistency of the DFT ensemble representation. 

The ensemble representation of Eq.~(\ref{eq:w-wts}) has an interesting physical interpretation. It is the weighted sum of \textit{all possible isolated-atom electronic structure problems}---neutral, cation, anion, and their ground and excited states---associated with a given atom of nuclear charge $Z$, {\it i.e.,} its atomic density ``fingerprint"; see Fig.~\ref{fig:Potentials}(c).  In practice, some states---for example most negative ions---will not contribute at all. But since the ensemble weights of the densities and corresponding energies self-adjust over the course of a dynamical simulation in response to system-wide chemical potential equalization, the states that do end up contributing will constitute a detailed quantum mechanical probe of the response of each atom's electrons to the evolving chemical environment. As a simulation proceeds, the contributions (weights) of the isolated atom densities associated with a given atom will evolve. This provides a natural mechanism for the potential to describe bond-breaking: the density ensembles corresponding to each bonded atom and their corresponding energy ensemble will reduce to a single, neutral, isolated-atom contributions.  

Another important feature of the ensemble representation is that since the atom-in-molecule densities are represented in terms of isolated atom states, the basis densities $\rho_{ijk}({\bf r})$ can computed in advance, and fitted with analytic forms for computational convenience.  This provides good control over the quantum mechanical representation, since atomic densities can be computed to high accuracy using standard quantum chemistry codes and a high-quality basis set.  In practice, the atomic densities can be computed on a 3D grid, sphericalized, and the resulting radial distribution functions fitted with sums of parameterized hydrogenic functions in an Aufbau-like procedure.\cite{amokwao2020,samuels2025,amokwao2026} Importantly, and in the spirit of using all available quantum mechanical information to construct an accurate latent space representation, two known asymptotic constraints are applied in the sphericalized density fits: the exponential Kato cusp at the nucleus,\cite{Kato1957,Steiner1963} and the known long-range exponential decay behavior of the atomic density.\cite{levy1976,Tal1978}  The sphericalization procedure ensures that the representation will scale linearly in the number of atoms, while describing the chemical environment without the need for explicit angular components. 

A rationale for the use of sphericalized densities can be found in recent theoretical results for spherical DFT,\cite{theophilou2018,nagy2018,samuels2026information} where it has been shown that the set of electron densities sphericalized about each atom in a larger Coulombic system encodes the same information about the electronic structure of the system as the total electron density in Hohenberg-Kohn DFT.  While the spherical DFT densities are manifestly not the same as atom-in-molecule densities,\cite{nagy2018,nagy2019} the recent distance geometry-based extension of spherical DFT\cite{samuels2026information} provides geometrical insight into the use of spherical densities for characterizing the local chemical environment of an atom.  Passaro and Zitnick's work on reducing the computational complexity of equivariant graph neural network embeddings using spherical channels,\cite{passaro2023reducing} which has been implemented in the UMA MLIP,\cite{wood2025family} is a potential connection point  between the ECT-EAM and equivariant message-passing GNNs that will be worth exploring.

\vspace*{.05in}
\noindent
3. {\it Atomic-state-dependent energy embedding functions.}\hfill \\
The embedding functions for neutral atoms in the original EAM were constructed empirically by fitting to experimentally-determined material properties.\cite{foiles1986,mishin2001} In the ECT-EAM, additional embedding functions are required for the ionic and excited atomic states participating in each atom's energy ensemble. Recent work\cite{baxter2026} has uncovered an unexpected three-way concordance between early numerical DFT results for the embedding functions of neutal atoms as a function of embedding density,\cite{puska1981} embedding functions computed based on the Hohenberg-Kohn DFT derivation of the EAM by Daw,\cite{daw1989} and the empirical $\rho \ln \rho$ embedding function proposed by Baskes as part of MEAM.\cite{baskes1987}  (See Fig.~2 of [\onlinecite{daw1989}] for an example of an embedding function with the characteristic shape.)

This observation provides a straightforward route to computing embedding functions for ionic and excited states of atoms by requiring a concordance between independently-computed embedding functions obtained via the Daw DFT approach, and the Baskes $\rho \ln \rho$ form.  The final concordance curve becomes the embedding function for the specified state of a given atom.\cite{baxter2026}  Remarkably, achieving parity between the two methods requires only minimal adjustments to the two parameters appearing in each of the models: an overall energy scale factor and reference density in the Baskes form, and prefactors for the local density approximation and von Weis\"{a}cker noninteracting kinetic energy terms in the Daw DFT expression. The final parameters remain close to physically-reasonable values in all cases.\cite{baxter2026} 

\vspace*{.05in}
\noindent
4. {\it Electrostatic interactions, computed via the $\rho_i^*({\bf r})$.} \hfill \\
As discussed in Section~\ref{sec:Background}, early MLIPs focused on the short-range interaction component of the potential. In generalizing to systems with long-range electrostatic interactions, the models were often modified to calculate electrostatic interactions based on effective charges computed from a harmonic model and a local, updatable model of the electronegativity.  In the ECT-EAM, the classical electrostatic interaction term is an intrinsic component of the model and its DFT derivation.\cite{daw1989} Accordingly, the ECT-EAM representation of the electrostatic interaction is expressed in terms of interacting ensemble atom-in-molecule densities, as seen in the second term of Eq.~(\ref{eq:ensFF}). To maintain internal consistency, the $\rho_i^*({\bf r})$ appearing in the argument to the state-dependent embedding functions (Eq.~(\ref{eq:AIM-dens})) and in the electrostatic interaction term are required to be the same for all atoms $i$.

\section{Discussion and conclusions\label{sec:Discussion}}
The constructive latent space approach presented here suggests a new direction for addressing the complex challenges inherent in interatomic potential design. The ECT-EAM ``folds in" detailed quantum mechanical information about an atom's local chemical environment, and can serve as a reference framework\cite{valone2006electron} for further theoretical and computational refinements. The deep pre-organization of bonding patterns based on the  ensemble picture avoids the curse of dimensionality bottleneck and the need to sample large and necessarily incomplete training datasets in order to identify a limited subset of bonding motifs. The latent space constructs are built upon a rigorous foundation of DFT, providing a natural link between the electronic (quantum mechanical) and atomistic (classical) length scales, enabling a principled description of charge excitation, charge polarization, charge transfer, and electron correlation. The foundation in DFT provides opportunities for further developments in the latent space representation, as new exchange-correlation density functionals and new formulations of ensemble DFT emerge.\cite{fromager2025ensemble} A key advantage is that since the latent variables are defined {\it a priori} in terms of physical constructs, models built upon its foundation, are {\it de facto} interpretable: reverse engineering for explainability is not required. 

The use of close-to-exact atomic basis densities to express the ensemble atom-in-molecule latent space patterns stands in contrast to the traditional use of Gaussian radial basis functions in most MLIPs. The atomic basis densities reflect native atomic electronic structure, with exponential (Slater-type) rather than Gaussian behavior. The electronic structure and densities of atoms can be computed essentially exactly, using standard quantum mechanical techniques. Fitting the densities as radial distribution functions\cite{amokwao2020,amokwao2026} enhances sensitivity to the shell structure of the atoms.  Imposing the Kato nuclear cusp \cite{Kato1957,Steiner1963} and long-range asymptotic exponential decay\cite{levy1976,Tal1978} constraints\cite{amokwao2026} ensures that both short- and long-range dynamical electron correlation will be described correctly in the ensemble density description of large systems.\cite{hollett2011two} 

There are a number of immediate directions to pursue in order to explore the accuracy and scalability of the constructive latent space approach. The  representation of the atom-in-molecule densities in terms of atomic states makes it straightforward to evaluate effective atomic charges based on the computed state weights of a given system configuration.  The use of effective charges to construct force fields for macromolecular simulation and interpret chemical interactions has a rich history, and it will be of interest to compare ECT-EAM-predicted charges with some of these empirical approaches, particularly the DFT-inspired topological approach of Bader.\cite{bader1985,bader1990}  To test the internal consistency of the model, 
density ensemble weights computed by fitting atomic densities to molecular densities\cite{amokwao2020,samuels2025} can be compared to independently-computed energy ensemble weights derived from fitting to potential energies.\cite{baxter2026} 

An important next step will be to implement the ECT-EAM potential as a force field for dynamical simulation. This will require a numerical strategy for global chemical potential equalization,\cite{Atlas2021} to allow the dynamical rebalancing of weights defining the ensembles of densities and energies, and maintain an energy-minimizing balance between local embedding and global electrostatic contributions. The net effect of this rebalancing can be understood as an instantaneous redefinition of the interaction potential that tracks with the evolving atomic configuration.  Since the ensembles include explicit contributions from excited atomic states, the latent space representation describes excited potential energy surfaces of the system as a whole, which may be sampled over the course of the dynamics, due to avoided crossings.\cite{yarkony2012nonadiabatic}  In a simple diatomic system, excited states are conventionally described in terms of their dissociation channels to excited atomic configurations (see, for example, Fig.~1 of Ref.~[\onlinecite{musial2014first}].)  For larger atomic systems, the analogous potential energy surface representation is not possible due to the high dimension of the configurational coordinate space. In the ensemble representation, however, the potential energy surface is fully characterized by its atomic basis function weights. This opens the possibility of exploring non-adiabatic molecular dynamics using ECT-EAM: surface hopping\cite{tully1990} without explicit hops.

The evolution of MLIPs in the direction of message-passing graph neural networks, with sophisticated geometric strategies for imposing equivariance,  provides an interesting opportunity to tune the relative balance between nodes and edges in describing the interaction physics. An approach such as the ensemble embedding framework could be used to shift the responsiblity for effective modeling of short-range physics decisively in the direction of local nodes (extreme ``nearsightedness"), leaving responsibility for perturbative corrections to the GNN's message passing layers. MACE is built upon the atomic cluster expansion (ACE),\cite{drautz2019atomic,dusson2022atomic} 
which was designed to incorporate atom-centered density-like features---in its initial formulation, denoted by $\rho_i^{(1)}$, $\rho_i^{(2)}$, and $\rho_i^{(3)}$---generalizing 
the MEAM angular-dependent density expansion.\cite{baskes1987,baskes1989,baskes1992modified}  A key feature of ACE is the use of these density-like features as arguments to a nonlinear ``embedding energy function."\cite{drautz2019atomic,dusson2022atomic}  The initial choice of nonlinear embedding was motivated by the Finnis-Sinclair $\sqrt{\rho}$ form, but other embedding functions, such as Baskes' $\rho \ln \rho$, concordant with Daw's DFT embedding function,\cite{baxter2026} may provide useful alternatives, particularly if the more general ensemble embedding construct is used. Independent of the precise functional form, the nonlinear embedding term serves to package the physics into an atomic (node-centric, in GNNs) form, to improve scaling to larger systems.\cite{drautz2019atomic}  
This highlights an unexpected and potentially deep connection between contemporary large-scale 
machine learning models and the latent space perspective, for discovering and predicting the properties of novel molecules and materials.

\section*{Acknowledgements} 
The author would like to express her gratitude to Steven Valone for many stimulating and enjoyable discussions and longstanding collaboration on the design of fragment- and atomic-excitation-based interatomic potentials, which have helped inspire the latent space interpretation presented here. 
The author also thanks Godwin Amo-Kwao, Chance Baxter, Jonas Dittman, Vijay Janardhanam, Elan Landau, Krishna Muralidharan, and Sol Samuels for their contributions to the development and testing of the ECT-EAM and latent space representation. This work was supported in part by the National Science Foundation under grant No.~CHE-0304710, and by the DoD/DTRA CB Basic Research Program under grant No.~HDTRA1-09-1-0018.  We are grateful to the UNM Center for Advanced Research Computing, supported in part by the National Science Foundation and NSF grant No.~OCI-1040530, for providing the computational resources used in this work. 

\bibliography{LatentSpace}

\end{document}